# TRANSPORT STUDIES OF LPA ELECTRON BEAM TOWARDS THE FEL AMPLIFICATION AT COXINEL

M Khojoyan, F. Briquez, M. Labat, A. Loulergue, O. Marcouillé, F. Marteau, G. Sharma,
M.E. Couprie, Synchrotron SOLEIL, France

*Abstract*

Laser Plasma Acceleration (LPA) [1] is an emerging concept enabling to generate electron beams with high energy, high peak current and small transverse emittance within a very short distance. The use of LPA can be applied to the Free Electron Laser (FEL) [2] case in order to investigate whether it is suitable for the light amplification in the undulator. However, capturing and guiding of such beams to the undulator is very challenging, because of the large divergence and high energy spread of the electron beams at the plasma exit, leading to large chromatic emittances.

A specific beam manipulation scheme was recently proposed for the COXINEL (Coherent X-ray source inferred from electrons accelerated by laser) setup, which makes an advantage from the intrinsically large chromatic emittance of such beams [3]. The electron beam transport is studied using two simulation codes: a SOLEIL in-house one and ASTRA [4]. The influence of the collective effects on the electron beam performance is also examined.

## INTRODUCTION

Laser-plasma based accelerators [5] are promising alternatives for the generation of high energy electron beams as they sustain orders of magnitude higher fields compared to the conventional accelerators. In the few hundreds of megaelectronvolt (MeV) range the produced electron beams typically have the following performance: a few kA peak current [6], a few percent energy spread [7, 8, 9] and transverse emittance comparable [10, 11] or even smaller [7, 12] than the emittance values achieved by the conventional linear accelerators (CLA) [13]. Indeed, the small transverse emittances correspond to submicron LPA transverse size and a few mrad [6, 11] beam divergence. The high divergence of LPA electron beam at the exit of plasma can be reduced (by about factor of two) by applying an additional plasma section as a transverse focusing element [14]. An adiabatic matching solution [15] may also help in minimizing the electron beam divergence and transverse emittance growth after the plasma. Even though, the LPA beam should be refocused by using ultrahigh gradient quadrupoles (requiring the permanent magnet technology) positioned very close to the electron source. One has also to cope with the large LPA beam energy spread either with an optimized undulator design using chicane decompression [16], or by a transverse-gradient undulator [17].

Still, the large divergence and high energy spread of such beams makes the capturing and the beam transportation very complicated for further applications such as FELs. Shortly after leaving the plasma, in the refocusing section, a rapid development of transverse position-energy correlation of the particles occurs within the electron bunch, which results in strong beam quality degradation by means of emittance worsening and bunch lengthening [18, 19, 20].

COXINEL [21] project within the frame of LUNEX5 [22, 23] aims at demonstrating FEL amplification using LPA. Recently a chromatic matching manipulation scheme was proposed for COXINEL which turns the large chromatic emittance of LPA beams into a direct FEL gain advantage. For instance, numerical simulations have shown up to two orders of magnitude further enhancement in the FEL peak power [3] over 5 m undulator.

This paper presents the electron beam transport studies at COXINEL, after the exit of plasma up to the undulator. LPA beam parameters and the simulation tools for the beam transport are first described. The electron beam dynamics is then examined disregarding the collective effects. The impact of the chicane decompression on the beam performance without and with the collective effects is studied as well.

## SETUP FOR LPA BEAM TRANSPORT

The setup of COXINEL is shown in Fig. 1. The electrons will be generated from a 60 TW laser provided by LOA (Laboratoire d'optique appliquée) and transported along the COXINEL transfer line.

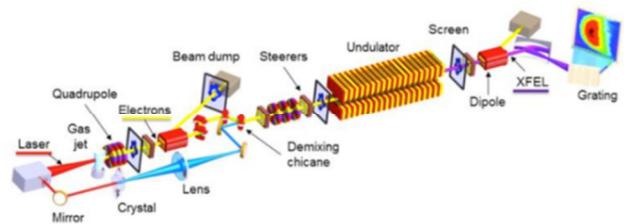

Figure 1: The layout of COXINEL for LPA based FEL. The electron beam transport line includes strong permanent magnet quadrupoles of variable gradient, located right after the plasma chamber, a demixing chicane for energy sorting of the electrons and a second set of conventional quadrupoles for the proper focusing of different electron slices inside the undulator. The distance from the plasma exit until the undulator center is 6.7 m.

The baseline parameters of LPA beam for COXINEL are shown in Tab. 1. They have been defined according to the best experimentally achieved beam performance at LOA [9, 14], meanwhile expecting some further progress from the LPA community. Indeed, the assumed value of the beam energy spread is still about two orders of magnitude larger as compared to the beams from CLA.

Table 1: Set of 180 MeV electron beam parameters for COXINEL studies.

| LPA beam parameter | Value |
|---|---|
| Charge, pC | 34 |
| Energy, MeV | 180 |
| Peak current, kA | 4 |
| Bunch rms length, µm | 1 |
| Relative rms energy spread, % | 1 |
| Transverse rms divergence, mrad | 1 |
| Normalized rms emittance, π mm mrad | 1 |

Three permanent magnet quadrupoles with adjustable strength (up to ~ 200 T/m) are located very close to the electron source to refocus the beam. A dipole chicane (of variable strength) consisting of four identical magnets follows afterwards for bunch decompression [16, 24]. Another set of electromagnetic quadrupoles (up to ~ 20 T/m) is used for dedicated beam matching, where the slice electron beam waist along the undulator is synchronized with the FEL wave propagation by adjusting the chicane strength [3]. There are other beamline components such as steerers, beam position monitors, screens, etc., for the electron and photon beam characterization [21].

The beam transport at COXINEL from the exit of plasma up to the undulator is investigated by cross-checking two different tracking softwares: SOLEIL in-house code and ASTRA [4]. First, BETA code [25] is used to match the beam optical functions at the undulator center based on the source to image (S2I) standard optics. The matching includes chromatic terms up to the second order. The in-house code simulates the beam transport by symplectic mapping [26] through each magnetic element (assuming hard edge magnetic field profile). The tracking includes collective effects such as 3D space charge (SC) and coherent synchrotron radiation (CSR) [27]. The second code (ASTRA) applies an integration method to solve the equation of motion of a charged particle in an electromagnetic field. The fields contain the contribution terms from external and the SC fields. For the SC field calculation, both codes compute the electrostatic potential in the rest frame of the bunch using the FFT (Fast Fourier Transform) Poisson solver [28]. As the in-house code tracks the particles based on element by element mapping, its calculation time is much faster compared to ASTRA.

## BEAM DYNAMICS STUDIES WITHOUT COLLECTIVE EFFECTS

Firstly, an electron beam behavior along the COXINEL beamline is investigated without any collective effects. A perfect electron beam of 6D Gaussian distribution without any correlations is presumed in the simulations. The optimum lattice configuration obtained by the in-house code has been put in ASTRA for comparison of the output results. The evolution of the normalized transverse emittance along the COXINEL line is presented in Fig. 2, comparing the in-house and ASTRA codes. Similar results from two different tracking codes are found. An increase of the total emittance due to the chromatic emittance [20] occurs at the first quadrupole triplet, where the beam divergence is still large. The emittance values after the triplet are fairly constant. In our case, the relative difference between the initial and the total emittances (chromatic term) is not significant but it becomes considerable for smaller values of the initial emittance.

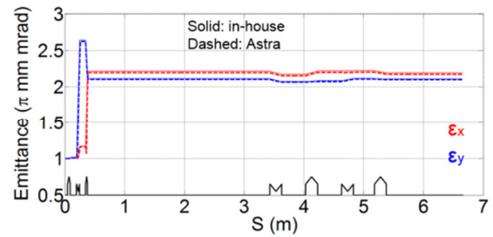

Figure 2. Trace-space normalized transverse emittances (red: horizontal and blue: vertical) along the beamline for ASTRA (dashed) and in-house (solid) code. Sketch of the focusing / defocusing magnets shown at the bottom of the figure. LPA beam parameters of Tab 1.

The evolution of the beam envelope (rms beam size) and rms bunch length along the COXINEL lattice, shown in Fig. 3, exhibit well matched curves for the two tracking codes. The sharp increase in bunch length inside the first three quadrupoles is due to the large divergence. Additional bunch lengthening after the triplet takes place due to different velocities of the electrons inside bunch.

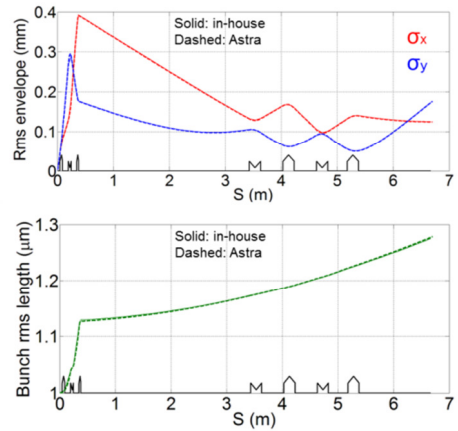

Figure 3. Evolution of the beam rms envelope (upper plot) and the bunch rms length (lower plot). Dashed line: ASTRA, solid line: in-house code.

The electron beam distribution in transverse (horizontal beam size and beam divergence) and longitudinal (internal bunch coordinate and relative energy spread) phase spaces is shown in Fig. 4 for two tracking codes: in-house (upper plots) and ASTRA (lower plots). The beam shape in the transverse phase space (left part of the figure) is the result of contribution of the chromatic terms. The beam distribution in the longitudinal phase space (right part of the figure) is obtained by taking into account the velocity difference of the particles with respect to their longitudinal coordinate. One can notice the high divergent particles, shifted to the left side of distribution.

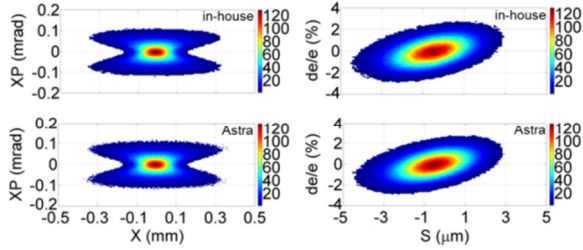

Figure 4. Electron beam distribution in transverse (left) and longitudinal (right) phase spaces from in-house code (up) and ASTRA (down).

Fig. 5 shows the slice beam properties at 6.7 m after the source without any collective effects. The trailing particles with high divergence are the reason for the large emittance values at the bunch tail inside the slice emittance distribution (Fig. 5a). The beam peak current (Fig. 5b) is decreased by about 25 % (initially being 4 kA) accumulating the effects of large divergence and high energy spread. From the right set of the plots the energy de-mixing effect of the chicane on the slice emittance (Fig. 5 d) distribution is visible. Moreover, in this case of the $r_{56} =1$ mm linear chicane strength, the beam slice energy spread is improved by an order of magnitude at the cost of the peak current.

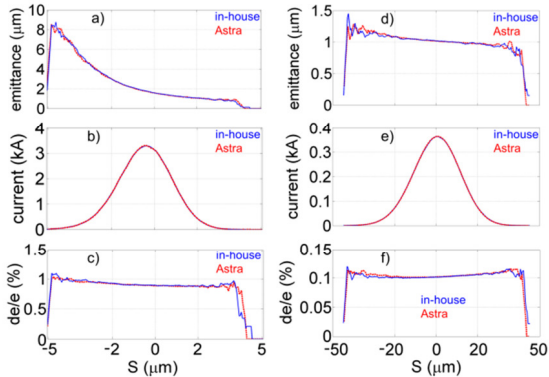

Figure 5. Horizontal slice emittance (a, d), slice current (b, e) and relative slice energy spread (c, f) at 6.7 m after the source without any collective effects. Blue (solid): in-house code, red (dashed) ASTRA. a, b, c: the case without chicane, d, e, f: the case with chicane. Horizontal axis is the internal bunch coordinate.

## BEAM DYNAMICS INCLUDING COLLECTIVE EFFECTS

As the electron beam has high peak current and very small transverse and longitudinal sizes after plasma (see Tab. 1), the SC effects can be essential during the beam propagation. Indeed, the chicane bunch decompression (~ 10 times bunch lengthening here) greatly relaxes the SC effects as can be seen from Fig.6, which is the analog of figure 5, including the SC effects. Strong impact of the SC effects on the beam performance is noticeable in the case without the chicane. The impact of the SC on the slice emittance distribution can be seen from Fig. 5d). The lower value of the peak current (figures 5e) and 6e)) is the result of additional bunch lengthening due to the SC. A very similar output of both tracking codes is also remarkable.

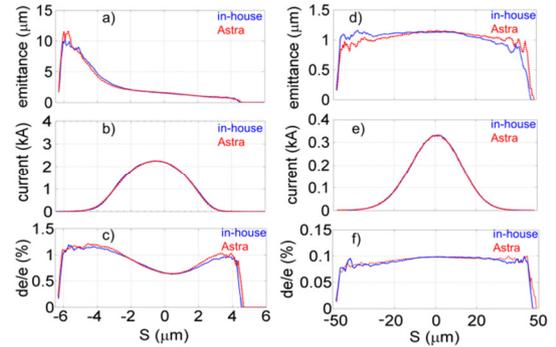

Figure 6. Electron beam slice properties at 6.7 m after the source including space charge effects. Blue (solid): in-house code, red (dashed) ASTRA. The left part: the case without chicane, the right part: the case with chicane.

Fig.7 depicts the slice beam properties obtained from the in-house code for three different cases: without collective effects, including the SC and 1D CSR effects in addition to the space charge. The results are obtained for the similar peak currents (by small adjustment of the chicane strength) including the effect of chicane de-mixing.

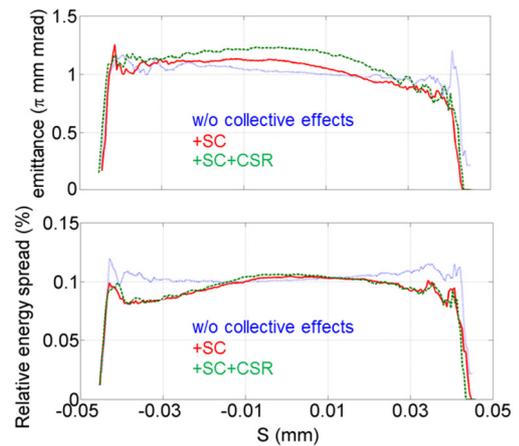

Figure 7. Electron beam slice emittance and slice energy spread at the undulator center (for ~ 350 A peak current value) calculated with the in-house code. Blue: without collective effects, red: space charge effects included, green: 1D CSR effects and SC included. LPA beam parameters of Tab 1.

About 10 % increase of the slice emittance (in the central part of the distribution) can be observed when the space charge effects are included. Another ~ 15 % increase in emittance occurs due to the CSR effects. The beam slice energy spread is slightly influenced (~ 5 %) by the mentioned effects. Changing the chicane strength would result in different contributions of different effects such as the 'chromatic emittance', the CSR and the electron beam-FEL synchronization effects. Particularly, for the case of lower initial beam emittance, the collective effects become essential. For instance, in the case of 0.1 mm mrad emittance (for the same machine configuration as for the case of 1 mm mrad emittance) the simulations show ~ 6 times dilution of the slice emittance after the beam propagation. The beam quality degradation is the result of the partial de-mixing effect of the chicane and the SC. This emphasizes, that having initially small emittance is not always beneficial for LPA-driven FEL.

## CONCLUSION

Electron beam transport has been studied at COXINEL from the plasma exit until the undulator. The large initial beam divergence was suppressed by the first quadruplet, the slice beam energy spread was reduced ~ 10 times by adjusting the chicane strength ($r_{56}$) to 1 mm value. The influence of space charge effects on the beam dynamics was investigated. It was found that chicane decompression significantly reduces the influence of the space charge on the LPA beam performance during the transport. In all the mentioned cases the beam output was obtained by cross-checking two tracking softwares. Very good agreement between the tracking tools was found for various cases. Additionally, the effects of the SC and the CSR on the beam slice properties were estimated at fixed chicane strength. About 10 % increased value in the central part of the slice emittance distribution was obtained due to the SC effects. Additional ~ 15 % emittance increase was found when the CSR effects are taken into account. The beam slice energy spread was found to be not much affected by the collective effects.

Beam dynamics studies should be continued assuming different initial beam parameters (e.g. higher energy of 400 MeV, lower emittance of 0.1 mm mrad). In that case scanning of the chicane strength will be necessary to find an optimum working condition among aforementioned effects. Specially, for the case of lower initial emittance, the relative difference between the initial and the total emittances becomes larger and can have significant impact on the beam performance during the transport. Sensitivity studies of the acceptance of the initial LPA beam parameters and the possible misalignment of magnetic elements on the FEL output are also very important. Additional sets of simulations (plasma, in-house code and FEL simulations) are necessary for more detailed overview of the FEL performance at COXINEL.

## ACKNOWLEDGMENTS
The work is supported by the European Research Council Advanced Grant COXINEL (340015).